\documentclass[aps,pra,twocolumn,showpacs,superscriptaddress]{revtex4}

\usepackage{graphicx}
\usepackage{amsmath}
\usepackage{amssymb}

\newcommand{\br}{{\bf r}}
\newcommand{\bk}{{\bf k}}

\begin{document}
\bibliographystyle{prsty}

\title{Helical spin textures in dipolar Bose--Einstein condensates}
\author{J.~A.~M.~Huhtam\"aki} \email{jam@fyslab.hut.fi}
\affiliation{Department of Applied Physics/COMP, Aalto
University, P.O. Box 14100, FI-00076 AALTO, Finland}
\affiliation{Department of Physics, Okayama University, Okayama
700-8530, Japan}

\author{P.~Kuopanportti}
\affiliation{Department of Applied Physics/COMP, Aalto
University, P.O. Box 14100, FI-00076 AALTO, Finland}

\begin{abstract}
We numerically study elongated helical spin textures in ferromagnetic spin-1
Bose--Einstein condensates subject to dipolar interparticle forces. Stationary states of the
Gross--Pitaevskii equation are solved
and analyzed for various values of the helical wave vector and
dipolar coupling strength. We find two helical spin textures which
differ by the nature of their topological defects. The spin structure hosting a pair of Mermin--Ho
vortices with opposite mass flows and aligned spin currents is
stabilized for a nonzero value of the helical wave vector.
\end{abstract}

\pacs{03.75.Lm, 03.75.Mn, 67.85.Fg, 67.85.Bc}

\keywords{Bose-Einstein condensate, dipole-dipole interaction, spin texture, helix}

\maketitle

\section{Introduction}
Helical structures lie at the heart of several feats of innovation.
The Archimedean screw and spiral staircases are inventions
dating back to ancient history. Simple bolts and spring coils stand as hallmarks
of practicality in modern everyday life. Helical growth is a clear demonstration
that such structures also exist naturally without human intervention.
On the other hand, helices are vital for life itself as we know it,
the DNA polymer being a famous example of a double-helix structure~\cite{Watson1953}.

The study of helical spin textures has recently drawn attention in the field of
gaseous spinor Bose--Einstein condensates (BECs). In a quantum degenerate gas of ferromagnetic spin-1 ${}^{87}{\rm Rb}$,
a helical magnetization texture was observed to decay into small spin domains~\cite{Vengalattore2008}.
This effect was argued to result from the weak interatomic magnetic dipole forces in the system.
Closely related to the experiment, the dynamical instability of an $XY$ spiral state has been investigated theoretically~\cite{Cherng2008}.

It has been predicted that a sufficiently strong magnetic dipolar forces can spontaneously give rise to
intriguing spin textures in ferromagnetic condensates: the long-range dipolar potential stabilizes
spin-vortex states in various geometries, as was demonstrated using a \mbox{spin-1} model~\cite{Yi2006,Kawaguchi2006}
and a classical spin approach~\cite{Takahashi2007,Huhtamaki2010}.
Even weak dipolar interactions can lead to dynamic formation
of a helical spin texture in a ferromagnetic spin-1 BEC~\cite{Kawaguchi2007}.
In the absence of external magnetic fields, the spin helix can appear as the ground-state texture in a suitable geometry and with strong enough dipolar
interactions~\cite{Huhtamaki2010}.

In the present article, we investigate helical spin textures
in ferromagnetic dipolar condensates using a spin-1 model. We find stationary helical
states under the assumption that the system is infinitely long in the direction of the helix axis and find solutions hosting different types
of topological line defects, i.e., quantized vortices.
Due to the symmetry of the order-parameter field, these line defects
encircle the condensate in a helical pattern. The resulting vortex structure resembles a pair of vortices
excited in Kelvin modes in a stationary configuration.
Direct experimental evidence of Kelvin waves has been observed in a spin-polarized BEC of ${}^{87}{\rm Rb}$
by studying two transverse quandrupole modes of the atomic cloud~\cite{Bretin2003}. Kelvin-wave excitations of quantized vortices
have also been studied theoretically in a similar setup~\cite{Mizushima2003,Fetter2004,Simula2008}. Moreover, helical vortices in a two-component
condensate have been investigated~\cite{Cho2005}.

Although we focus on a spin-1 BEC, the study can also shed light on phenomena in strongly dipolar systems with more complex order parameters.
The energy-minimizing helical spin textures that we obtain are in qualitative agreement with those found earlier using the classical
spin approach~\cite{Huhtamaki2010}. Since
the classical spin model is supposed to be accurate for ferromagnetic systems in the limit of large magnetic moments, it is plausible to expect that similar
structures exist independent of the value of the atomic spin number $F$.
The most timely example of a condensate with larger $F$ and significant dipolar interactions is the spin-3 gas of ${}^{52} {\rm Cr}$~\cite{Griesmaier2005} which
has been produced by purely optical means~\cite{Beaufils2008}. The chromium atoms have magnetic moments of $6\, \mu_\mathrm{B}$, whereas
the maximal atomic magnetic moment for an alkali-metal condensate is $1\, \mu_\mathrm{B}$. Recently, there has been progress in cooling and trapping
vapors of $\mathrm{Tm}$ $(4\, \mu_\mathrm{B})$~\cite{Sukachev2010}, $\mathrm{Er}$ $(7\, \mu_\mathrm{B})$~\cite{Berglund2008},
and $\mathrm{Dy}$ $(10\, \mu_\mathrm{B})$~\cite{Lu2010}, the last having the largest atomic magnetic moment of all known elements.

\section{Theory}
In this work, we study ferromagnetic spin-1 BECs, such as
${}^{87}{\rm Rb}$, using a zero-temperature mean-field model. In
addition to the density--density and spin--spin interatomic forces,
we also include long-range dipolar interactions in the model.

Stationary states of the system are solutions to the
time-independent Gross--Pitaevskii (GP) equation
\begin{eqnarray}
\label{GP} \hat{h} \Psi(\br) &+& g n(\br) \Psi(\br) + g_{\mathrm s} \sum_\alpha
M_\alpha(\br) \hat{S}_\alpha \Psi(\br) \\
&+&g_{\mathrm d} \sum_{\alpha,\beta} \int d^3r' D_{\alpha \beta}(\br-\br')
M_\alpha(\br') \hat{S}_\beta \Psi(\br) = 0, \nonumber
\end{eqnarray}
where $\Psi=\left(\psi_1, \psi_0, \psi_{-1} \right)^T$ is a three-component spinor order parameter,
$\hat{h}=-\hbar^2 \nabla^2/2m+V_{\rm trap}(\br)-\mu$ is the
single-particle Hamiltonian, and $\hat{S}_\alpha$ denotes the
$\alpha$th component of the dimensionless $F=1$ spin operator
whose spin-space expectation value gives the $\alpha$th component of magnetization,
$M_\alpha(\br)=\Psi^\dagger(\br) \hat{S}_\alpha \Psi(\br)$. The density of particles is given by
$n(\br)=\sum \psi_k^*(\br) \psi_k(\br)$, where the components
$\left(\psi_1, \psi_0, \psi_{-1} \right)$ are the projections of
$\Psi$ onto the eigenbasis of $\hat{S}_z$. The total number of
particles per unit length, $\int d^2 r\, n(\br)=N$, is controlled through the chemical
potential $\mu$ acting as a Lagrange multiplier.

The coupling constants $g$, $g_{\mathrm s}$ and $g_{\mathrm d}$
measure the strengths of the local density--density, local
spin--spin, and non-local magnetic dipole--dipole interactions,
respectively. The first two are related to the scattering lengths $a_0$
and $a_2$ into spin channels with total spin $0$ and $2 \hbar$
through $g=4\pi \hbar^2 (a_0+2a_2)/3m$ and $g_{\mathrm s}=4\pi
\hbar^2 (a_2-a_0)/3m$. Throughout the work, we use 
$g_{\mathrm s}=-0.01\,g$, which is roughly the coupling
constant for ${}^{87}{\rm Rb}$
\cite{Klausen2001,Kempen2002,Widera2006} and a value previously used,
e.g., in \cite{Yi2006,Simula2010}. The dipolar coupling constant is
given by $g_{\mathrm d}=\mu_0 \mu_{\mathrm B}^2 g_{\mathrm
F}^2/4\pi$ with $\mu_0$, $\mu_{\mathrm B}$, and $g_{\mathrm F}$
being the permeability of vacuum, the Bohr magneton, and the Land\'e
factor, respectively. Rather than fixing $g_{\mathrm d}$ to some
particular value, e.g., $g_{\mathrm d} \sim 10^{-3} g$ as for
${}^{87}{\rm Rb}$, we present results for various interaction
strengths in order to emphasize the role of dipolar effects. Moreover,
by using greater values of $g_\mathrm{d}/g$,
the study should also provide useful information for systems subject to
strong dipolar forces, such as gases of ${}^{52}{\rm Cr}$
\cite{Lahaye2009}. In experiments, the ratio $g_\mathrm{d}/g$ may be controlled
with an optical Feshbach resonance~\cite{Fedichev1996}, which has been
demonstrated for ${}^{87}{\rm Rb}$~\cite{Theis2004}.

The long-range dipolar interactions are characterized by the
functions $D_{\alpha \beta}({\bf R})=\left(\delta_{\alpha \beta}
R^2-3 R_\alpha R_\beta \right)/R^5$, where $\{ R_\alpha\}$ denote
the components of the argument ${\bf R}$ and $R=\sqrt{\sum_\alpha R_\alpha^2}$.
The elements of the traceless symmetric tensor $D_{\alpha \beta}$
have a $d$-wave symmetric form and are thus simply expressed in
cylindrical coordinates. The interaction integral in Eq.~(\ref{GP})
can be viewed as an effective potential for the order parameter
$\Psi$ at $\br$ arising from the magnetization throughout the
system.

In the present work, we concentrate on elongated helical solutions
to Eq.~(\ref{GP}), similar to the helical textures studied recently
in cigar-shaped systems using a classical spin
approximation~\cite{Huhtamaki2010}. For simplicity, we assume an
infinitely long system in the axial direction, confined radially by
a harmonic potential $V_{\rm trap}=\frac{1}{2}m\omega_r^2 r^2$, where
$r=\sqrt{x^2+y^2}$ and $\omega_r$ is the radial trapping frequency.
The axial symmetry of the trap allows a well-defined wave vector
$\kappa$ for the helical texture. The aim is to fix $\kappa$ and
calculate the energy-minimizing texture for a given radial plane,
say, for $z=0$, subject to the condition that the planar texture is
mapped along the axial direction according to the helical structure.
Therefore, we write the order parameter in the general form
\begin{equation}
\label{ansatz} \Psi(r,\varphi,z)=e^{i q_z z}e^{i \kappa z
\hat{S}_z}\Psi(r,\varphi+\kappa z,0)
\end{equation}
expressed in cylindrical coordinates $(r,\varphi,z)$. The second
exponential factor describes a spin-rotation of angle $\kappa z$
about the $z$-axis in the clockwise direction.
Due to the non-linear terms in Eq.~(\ref{GP}),
the order parameter feels an effective potential with a period of
$2\pi/\kappa$ in the axial direction, and thus the Ansatz is written
in the form of a Bloch-wave including the first exponential factor.

By substituting Eq.~(\ref{ansatz}) into Eq.~(\ref{GP}) and
setting $z=0$, we obtain a GP equation reduced to
polar coordinates $(r,\varphi)$,
\begin{eqnarray}
\label{reduced_GP} \hat{h}' \Psi'(r,\varphi) &+& g n'(r,\varphi) \Psi'(r,\varphi) + g_{\mathrm s} \sum_\alpha
M_\alpha'(r,\varphi) \hat{S}_\alpha \Psi'(r,\varphi) \nonumber \\
&+&g_{\mathrm d} \sum_\beta I_\beta(r,\varphi,\kappa) \hat{S}_\beta \Psi'(r,\varphi) = 0,
\end{eqnarray}
where the primes denote that the quantities are evaluated at $z=0$. The kinetic
energy in the transformed single-particle operator $\hat{h}'$ is
given by
\begin{equation}
\label{kinetic_energy} -\frac{\hbar^2}{2m}\nabla^2 +
\frac{\hbar^2}{2m}\left[q_z + \kappa \left( \hat{L}_z + \hat{S}_z
\right) \right]^2,
\end{equation}
where $\nabla^2=\partial_r^2+\partial_r/r+\partial_\varphi^2/r^2$
and $\hat{L}_z=-i\partial_\varphi$. The functions
$I_\beta(r,\varphi,\kappa)$ arising from the dipolar interactions
are evaluated in Appendix A.

Only the kinetic-energy term in the energy functional depends
explicitly on the axial wave vector $q_z$, and hence minimization of
the total energy with respect to $q_z$ yields
\begin{equation}
q_z^{\rm min} = -\frac{\kappa}{N} \int d^2 r\, \Psi^\dagger \left( \hat{L}_z +
\hat{S}_z \right) \Psi.
\end{equation}
Substitution back into Eq.~(\ref{kinetic_energy}) reveals that for
a fixed planar texture, the kinetic energy is a quadratic function
of the helix wave vector $\kappa$. Moreover, the kinetic energy
becomes independent of $\kappa$ exactly when ${\rm Var} \big(
\hat{L}_z + \hat{S}_z \big)=0$, implying that $\Psi$ must be an
eigenstate of $\hat{L}_z + \hat{S}_z$. Denoting the (integer)
eigenvalue by $\lambda$, we find that in such a case the order
parameter is of the form
$\psi_k(r,\varphi,z)=e^{i(\lambda-k)\varphi}\psi_k(r,0,z)$, which is an
integer-spin vortex state. This describes the general form of states
for which the magnetization is invariant with respect to rotations
about the $z$-axis, implying that the dipolar energy and hence also
the total energy are independent of $\kappa$.

\begin{figure}[!t]
\includegraphics[width=180pt]{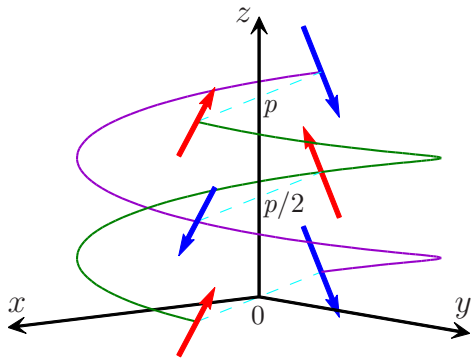}
\caption{\label{FIG1} (Color online) A schematic illustration of a
single period of a typical spin helix state. Here, $p=2\pi/\kappa$ denotes
the pitch of the helix whose wave vector is $\kappa$. The solid curves depict the helical trajectories
of two points on the $z=0$ plane, and the arrows point along the local direction of magnetization. If the
state hosts vortices, their cores will also follow similar trajectories.}
\end{figure}

Previous studies have shown that for very elongated condensates with
finite dipolar interaction strengths, $g_{\mathrm d} > 0$, the spin-polarized
textures with the magnetization lying along the long axis are the
energetically favored ones~\cite{Yi2006,Huhtamaki2010}. Such
states are of the general form discussed above and hence do not
depend on $\kappa$. In this study, however, we are interested in
helical textures which are found by requiring the symmetry
conditions~\cite{Huhtamaki2010}
\begin{eqnarray}
M'_{x,y}(r,\varphi+\pi) = M'_{x,y}(r,\varphi), \nonumber \\
\label{Symmetry_Conditions} M'_z(r,\varphi+\pi) = -M'_z(r,\varphi).
\end{eqnarray}
Equations~(\ref{Symmetry_Conditions}) are satisfied, e.g., when
$\psi_{k}(r,\varphi+\pi,z)=\psi^*_{-k}(r,\varphi,z)$. The textures
presented in the next section are solutions to Eq.~(\ref{GP})
subject to these symmetry conditions with respect to inversion
about the $z$-axis. A typical spin helix is shown schematically in Fig.~\ref{FIG1}.

\section{Results}

We have solved the reduced GP equation, Eq.~(\ref{reduced_GP}), for
various values of the helical wave vector $\kappa$ and the dipolar
coupling constant $g_{\mathrm d}$. In the numerical simulations, we
choose the harmonic oscillator length $a_r=\sqrt{\hbar/m\omega_r}$
for the unit of length and $\hbar \omega_r$ for the unit of energy.
For definiteness, we fix the value of the
dimensionless density--density coupling constant to $\tilde{g}=g N /\hbar
\omega_r a_r^2=10^3$, which corresponds roughly to $1.5 \times 10^4$
${}^{87}{\rm Rb}$ ${\rm atoms}/\mu {\rm m}$ in a trap with $a_r=1\,
\mu {\rm m}$.

\begin{figure}[!t]
\includegraphics[width=240pt]{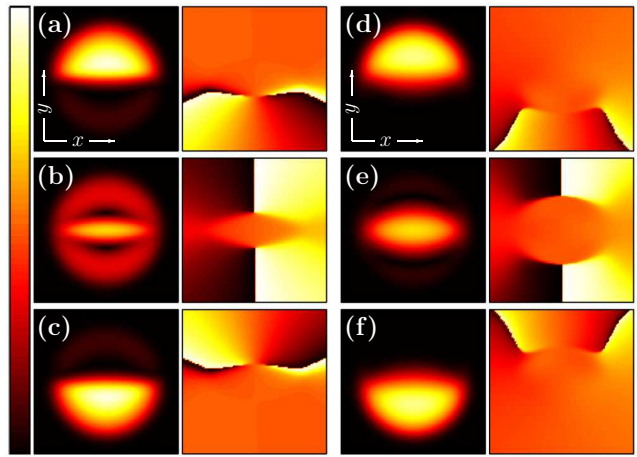}
\caption{\label{FIG2} (Color online) Illustration of the MHV
helix state for $g_{\mathrm d}=0.10\, g$ with $\kappa=0$ in (a)--(c)
and $\kappa=0.25\, {\rm rad}/a_r$ in (d)--(f). For both states, the
amplitudes of the order-parameter components ($\psi_1$,
$\psi_0$, $\psi_{-1}$) are given in the left column and
the complex phases in the right column. This class of
states hosts a pair of MHVs with aligned spin
currents and opposite mass flows. For each state separately, the
color-map range in the left column is $\big[0, {\rm max}
\left\{n(\br)\right\} \big]$ and in right column $\big[0,
2\pi\big]$. The field of view in each panel is $[14\, a_r \times
14\, a_r]$.}
\end{figure}

One particular class of stationary solutions is illustrated in
Fig.~\ref{FIG2} with $g_{\mathrm d}=0.10\, g$. On the left-hand side,
Figs.~\ref{FIG2}(a)--\ref{FIG2}(c), the state is shown in the
axially homogeneous case, $\kappa=0$, and on the right-hand side,
Figs.~\ref{FIG2}(d)--\ref{FIG2}(f), with finite helical pitch,
$\kappa=0.25\, {\rm rad}/a_r$. The left column in each
subfigure shows the amplitudes of the order-parameter components
$|\psi_1|$, $|\psi_0|$, and $|\psi_{-1}|$, from up to down. The complex
phases of the corresponding components are given in the right
column in each subfigure. For both states, the magnetization is
pointing predominantly in the positive $z$ direction in the upper
half $(y>0)$ and in the negative $z$ direction in the lower half
$(y<0)$ of the $xy$ plane.

The complex-phase plots in Figs.~\ref{FIG2}(a)--\ref{FIG2}(c) reveal
that the system is hosting two integer-spin vortices with phase
windings $(0,1,2)$ (upper half) and $(-2,-1,0)$ (lower half) in the
components ($\psi_1$, $\psi_0$, $\psi_{-1}$), respectively. The
ferromagnetic cores of the vortices are deformed into elliptic
shapes, which is also indicated by the separation of the phase
singularities in the $\psi_{\pm 1}$ components, cf.
Ref.~\cite{Simula2010}. The two vortices carry both spin- and mass
currents: the spin currents flow in the same direction whereas the
mass currents flow in the opposite directions, canceling the total
mass current in the state.

For finite helical wave vector, $\kappa>0$, the vortices move closer
to the surface of the cloud, as illustrated in
Figs.~\ref{FIG2}(d)--\ref{FIG2}(f). Not only do the cores of the
vortex pair separate farther apart for larger $\kappa$, but also the
separation between the singularities in the $\psi_{\pm 1}$ components
increases. However, one should note that for larger $\kappa$, the
vortex lines are tilted steeper with respect to the axial direction.
Due to the structure of the topological defects hosted by the order
parameter, we refer to this state as the Mermin--Ho vortex (MHV) helix.
This type of solution is found to exist in the
whole stability range of the system, $0 \le g_{\mathrm d}/g \lesssim 0.24$.

\begin{figure}[!t]
\includegraphics[width=240pt]{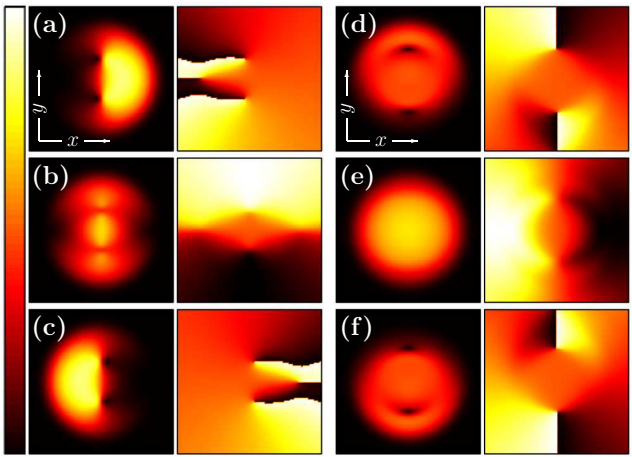}
\caption{\label{FIG3} (Color online) Illustration of the PCV
helix state for $g_{\mathrm d}=0.10\, g$ with $\kappa=0$ in (a)--(c)
and $\kappa=1.0\, {\rm rad}/a_r$ in (d)--(f). For both states, the
amplitudes of the order-parameter components ($\psi_1$,
$\psi_0$, $\psi_{-1}$) are given in the left column and
the complex phases in the right column. This class of
states hosts a pair of PCVs with aligned spin
currents and vanishing mass flows. For each state separately, the
color-map range in the left column is $\big[0, {\rm max}
\left\{n(\br)\right\} \big]$ and in right column $\big[0,
2\pi\big]$. The field of view in each panel is $[14\, a_r \times
14\, a_r]$.}
\end{figure}

The MHV helix is the ground-state texture within the symmetry requirements of
Eqs.~(\ref{Symmetry_Conditions}) for all parameter values considered
in this work. However, there exists an interesting class of excited
states which is illustrated in Fig.~\ref{FIG3} for $g_{\mathrm d}=0.10\, g$
with $\kappa=0$ and $\kappa=1.0\, {\rm rad}/a_r$ on the left- and
right-hand side, respectively. The complex-phase plots of
Figs.~\ref{FIG3}(a) and \ref{FIG3}(c) reveal that the state hosts
two spin vortices with the same phase windings $(-1,0,1)$ in the components
$(\psi_{1}, \psi_{0}, \psi_{-1})$, respectively. These
defects are spin vortices with polar (nonmagnetized) core regions.
Such vortices carry a spin current, whereas the mass current about the
vortex line vanishes. The polar vortex cores within the
ferromagnetic cloud are energetically analogous to air bubbles in
water, hence increasing the total energy of the system through
spin--spin interactions. The third singularity visible in
Figs.~\ref{FIG3}(a) and \ref{FIG3}(c) shows that the state hosts
also a pair of fractional half-quantum vortices close to the
surface of the cloud. However, these defects have very little effect
on the texture because the related phase gradients lie in regions of
nearly vanishing component amplitude $|\psi_{\pm 1}|$ for all values of
$\kappa$.

Solutions with similar vortex structures also exist for finite
helical wave vectors, as depicted in
Figs.~\ref{FIG3}(d)--\ref{FIG3}(f) for $\kappa=1.0\, {\rm rad}/a_r$.
Thus, we refer to this state as the polar-core vortex (PCV) helix. This
class of stationary states is found within the dipolar stability
range of the system, except for weak dipolar interaction strengths,
$g_{\mathrm d}/g \lesssim 10^{-3}$, likely due to numerical difficulties.

If one considers the vortex pairs in Figs.~\ref{FIG2} and \ref{FIG3}
as single entities, the total phase windings in both states are
$(-2,0,2)$ in the order-parameter components $(\psi_1, \psi_0,
\psi_{-1})$, respectively. By tracing the order parameter about the
$z$-axis along a path close to the surface of the condensate, the expectation
value of the spin rotates a total angle of $4\pi$ about the axial
direction. This suggests that both states represent doubly quantized
spin vortices that have split into different kinds of singly quantized
spin defects. The splitting of doubly quantized mass vortices in scalar
condensates has been studied previously both experimentally and
numerically~\cite{Shin2004,Mottonen2003,Huhtamaki2006,Mateo2006}.

\begin{figure}[!t]
\includegraphics[width=245pt]{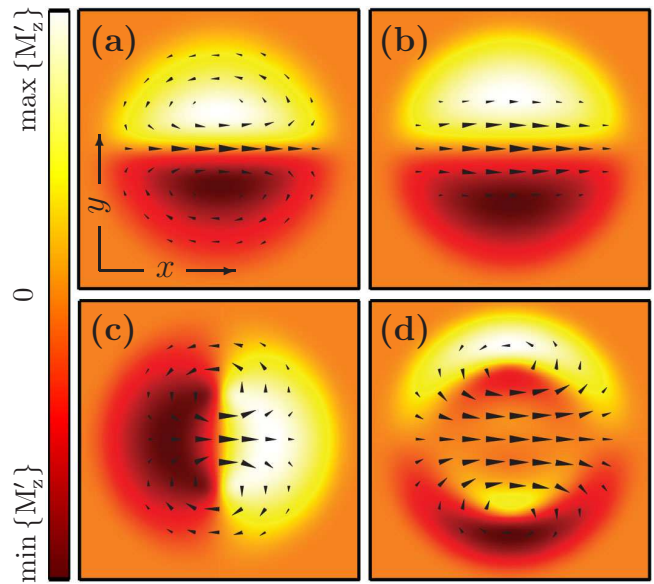}
\caption{\label{FIG4} (Color online) Illustration of the spin
textures for the states shown in Figs.~\ref{FIG2} and \ref{FIG3}.
Panels (a) and (b) correspond to the MHV helix state with
$\kappa=0$ and $\kappa=0.25\, {\rm rad}/a_r$, respectively. Panels (c) and
(d) refer to the PCV helix state with $\kappa=0$ and
$\kappa=1.0\, {\rm rad}/a_r$, respectively. The cones show the
projection of magnetization onto the $xy$ plane and the color
indicates the local value of axial magnetization $M'_z(x,y)$.
The field of view in each panel is $[13\, a_r \times 13\, a_r]$.}
\end{figure}

The spin textures of the states depicted in Figs.~\ref{FIG2} and
\ref{FIG3} are shown in Fig.~\ref{FIG4}. Figures \ref{FIG4}(a) and
\ref{FIG4}(b) illustrate the planar magnetization ${\bf M}'(r,\varphi)$ for
the MHV helix and
Figs.~\ref{FIG4}(c) and \ref{FIG4}(d) for the PCV
helix. The projection of magnetization on the $xy$ plane, ${\bf
M}'_{xy}(r,\varphi)=M'_x(r,\varphi) {\bf e}_x + M'_y(r,\varphi) {\bf e}_y$, is depicted
by the cones, with the length of each cone being proportional to the
local magnitude $\sqrt{M'^2_x(r,\varphi)+M'^2_y(r,\varphi)}$. The axial
magnetization, $M'_z(r,\varphi)$, is shown by color. At the center of the
system in the axially homogeneous textures ($\kappa=0$) in
Figs.~\ref{FIG4}(a) and \ref{FIG4}(c), the magnetization is pointing
perpendicular to $\nabla M'_z(r,\varphi)$ in the MHV and parallel to $\nabla M'_z(r,\varphi)$ in the
PCV state. For the MHV helix, this relative orientation is
maintained for all values of the helical wave vector $\kappa$.
However, for the PCV helix, the relative orientation
twists continuously with increasing $\kappa$ and finally locks into
the symmetric configuration shown in Fig.~\ref{FIG4}(d) for $\kappa
\sim 1\, {\rm rad}/a_r$.

\begin{figure}[!t]
\includegraphics[width=250pt]{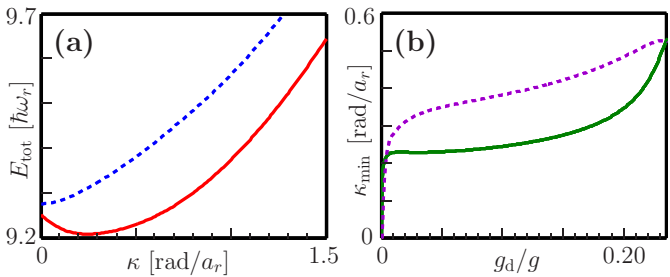}
\caption{\label{FIG5} (Color online) (a) Total energy per particle,
$E_{\rm tot}$, as a function of the helical wave vector $\kappa$ for
the MHV helix (solid curve) and the PCV helix (dashed
curve) with $\tilde{g}=1000$ and $g_{\mathrm d}/g=0.10$. The total
energy of the MHV helix exhibits a local minimum with
a finite wave vector $\kappa_{\rm min}$. (b) The energy-minimizing
wave vector $\kappa_{\rm min}$ for the MHV helix as a
function of the dipolar interaction strength $g_{\mathrm d}/g$ for
two values of the density--density coupling constant, $\tilde{g}=1000$ (solid
curve) and $\tilde{g}=200$ (dashed curve).}
\end{figure}

In Fig.~\ref{FIG5}, the total energy per particle, $E_{\rm tot}$, is shown as a function of the
helical wave vector $\kappa$ for the
MHV and the PCV helices. As previously, the coupling constants have the
values $\tilde{g}=1000$ and $g_{\mathrm d}/g=0.10$ for both states. The MHV helix
is the ground-state texture, within the constraints of
Eq.~(\ref{Symmetry_Conditions}), for all values of $\kappa$. Whereas
$E_{\rm tot}$ for PCV helix is minimized for the
axially homogeneous state, $\kappa=0$, the MHV helix
is stabilized for a finite wave vector $\kappa_{\rm min}$. This
minimum persists in the total energy for all positive interaction
strengths $g$ and $g_{\mathrm d}$ considered. However, the relative
magnitude of the dip, $\left[E_{\rm tot}(0)-E_{\rm
tot}(\kappa_{\rm min})\right]/E_{\rm tot}(0)$, decreases for weaker dipolar
interaction strengths. The curve $E_{\rm tot}(\kappa)$ is
roughly a shifted parabola, closely resembling the result
that long-period helical structures in MnSi and FeGe can become
stable due to ferromagnetic Dzyaloshinskii
instability~\cite{Bak1980}.

The energy-minimizing helical wave vector $\kappa_{\rm min}$ is
shown in Fig.~\ref{FIG5}(b) as a function of the dipolar interaction
strength $g_{\mathrm d}/g$ for two values of the dimensionless
coupling constant, $\tilde{g}=1000$ (solid curve) and $\tilde{g}=200$ (dashed curve).
In the absence of dipolar interactions, the total energy is
minimized by the axially homogeneous texture for which the kinetic
energy is minimal. The value of $\kappa_{\rm min}$ increases rapidly
as a function of $g_{\mathrm d}/g$ because the dipolar interactions,
which overwhelm the kinetic energy already for $g_{\mathrm d}/g
\approx 0.01$, favor a finite helical pitch. In general,
$\kappa_{\rm min}$ increases for stronger dipolar couplings. However,
the pitch of the energy-minimizing helix, $p_{\rm
min}=2\pi/\kappa_{\rm min}$, is not determined directly by the
dipole--dipole coherence length $\xi_{\rm dd}=\hbar/\sqrt{2m
g_{\mathrm d} n(0)}$, where $n(0)$ is the particle density at the
trap center. In fact, the coherence length $\xi_{\rm dd}$ shrinks
for larger particle numbers, whereas $p_{\rm min}$ typically
increases. This is likely due to the overall expansion of the condensate.

\begin{figure}[!t]
\includegraphics[width=250pt]{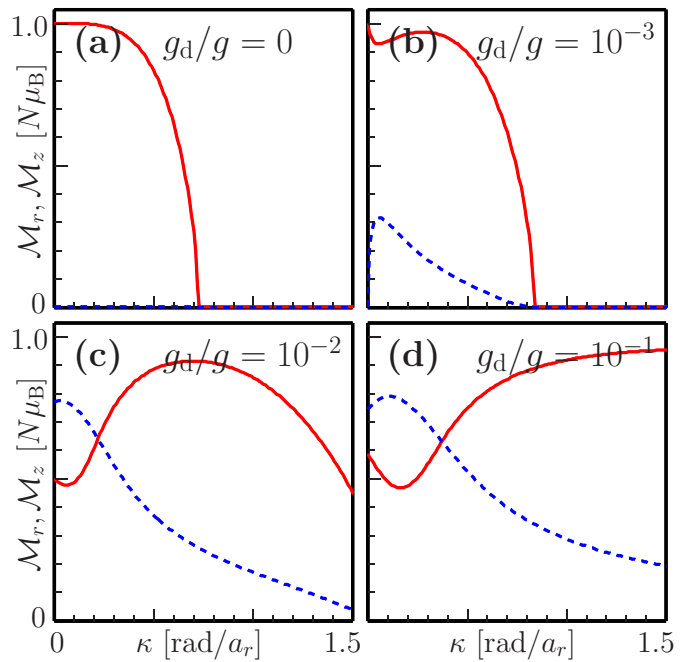}
\caption{\label{FIG6} (Color online) Integrated transversal $\mathcal{M}_r$ (solid curve) and axial $\mathcal{M}_z$ (dashed curve)
magnetization per unit length as a function of the helical wave vector
$\kappa$ for the MHV helix. The strength of the
dipolar coupling $g_{\mathrm d}/g$ is indicated in each 
panel. The overall magnetization is reduced for larger values of
$\kappa$, and the axial magnetization tends to increase with $g_\mathrm{d}/g$.}
\end{figure}

As a measure of how large the axial magnetization $M_z(\br)$ is on
average, we define the integrated axial magnetization per unit length as
$\mathcal{M}_z=\int d^2r |M'_z(r,\varphi)|$. Similarly, the integrated
transversal magnetization is given by $\mathcal{M}_r=\int d^2r
\sqrt{M'^2_x(r,\varphi)+ M'^2_y(r,\varphi)}$. These quantities are plotted in
Figs.~\ref{FIG6}(a)--\ref{FIG6}(d) as functions of the helical
wave vector $\kappa$ for $\tilde{g}=1000$ and four different values of the
dipolar coupling constant $g_{\mathrm d}/g$. The solid curve
corresponds to $\mathcal{M}_r$ and the dashed curve to $\mathcal{M}_z$.
As shown in Fig.~\ref{FIG6}(a), the axial magnetization
vanishes when $g_{\mathrm d}/g=0$. For $\kappa \gtrsim
0.7\, {\rm rad}/a_r$, the system becomes nonmagnetized because the
additional kinetic energy due to finite $\kappa$ for magnetized states exceeds
the energy gain from the spin--spin interaction, cf.
Ref.~\cite{Cherng2008}. Already for weak dipolar coupling,
$g_{\mathrm d}/g=10^{-3}$, which is roughly the value for
${}^{87}{\rm Rb}$, the axial magnetization becomes significant, as
shown in Fig.~\ref{FIG6}(b). Also, for finite dipolar coupling, the
value of $\kappa$ at which the system enters a polar state is
increased. The axial component of magnetization becomes larger for increasing $g_\mathrm{d}$, as illustrated in
Figs.~\ref{FIG6}(c) and \ref{FIG6}(d).

\section{Discussion}
In summary, we have studied helical spin textures in a spin-1
BEC subject to long-range dipolar
interactions. The axially elongated system was assumed to be confined radially by a
cylindrically symmetric harmonic potential. By using a helical
Ansatz, we reduced the zero-temperature GP equation
to a two-dimensional problem, in which the helical wave vector $\kappa$
appeared as a parameter. This allowed us to 
investigate states with variable values of $\kappa$. We found
two classes of helical solutions which we refer to as the Mermin--Ho
and the polar-core vortex helices, according to the structure of their
topological defects. Whereas the total energy
of the polar-core vortex helix is minimized for the axially
homogeneous case, $\kappa=0$, the Mermin--Ho vortex helix is
stabilized for a finite pitch.

The helical spin textures studied in this work are naturally most
transparent in elongated systems, such as cigar-shaped
BECs. One difficulty in observing them
as stable configurations in a condensate subject to weak dipolar interactions
is that the spins tend to align predominantly
parallel to the weak axis of the trap due to the head-to-tail
attraction of the dipoles~\cite{Yi2006,Takahashi2007,Huhtamaki2010}.
However, we have performed additional three-dimensional simulations indicating that this problem can be overcome
by placing the cigar-shaped system in a one-dimensional optical
lattice potential. A strong enough lattice deforms the elongated condensate into a series of
oblate clouds. Within each cloud, the preferred direction of
magnetization lies perpendicular to the weak axis of the trap, and
the relative orientation of magnetization between neighboring
systems is determined by the long-range part of the interaction
potential, i.e., by dipolar forces.

In the three-dimensional simulations, we calculated the total energy of an axially spin-polarized state
and a helical spin texture subject to the symmetry constraint in
Eq.~(\ref{Symmetry_Conditions}) as a function of the strength of the
optical lattice potential. We chose the aspect ratio of the
confining harmonic potential such that $\omega_z/\omega_r = 0.10$,
where $\omega_z$ and $\omega_r$ are the axial and radial trapping
frequencies, respectively. The values of the dimensionless coupling
constants were fixed to $\tilde{g}=5 \times 10^4$, $g_{\mathrm
s}/g=-10^{-2}$, and $g_{\mathrm d}/g=10^{-3}$, corresponding to $8 \times 10^5$ ${}^{87}{\rm Rb}$ atoms in a trap with $\omega_r=2\pi \times 100\, {\rm Hz}$. The optical
lattice potential was of the form $V_{\rm opt}=E_{\mathrm s} \sin^2
\left( k_{\mathrm s} z \right)$, where the constant $E_{\mathrm s}$
is the strength of the lattice potential, and the wave vector
$k_{\mathrm s}$ determining the distance between neighboring clouds
was fixed to $k_{\mathrm s}=\pi/10\, a_r$. The energy difference
between the helical and polarized states was found to decrease monotonously as a
function of $E_{\mathrm s}$, reaching degeneracy at
$E_{\mathrm s} \approx 17\, \hbar\omega_r$.

\begin{acknowledgments}
We acknowledge financial support from Japan Society for the Promotion of Science
(JSPS), the Emil Aaltonen Foundation, the V\"ais\"al\"a Foundation,
and Finnish Academy of Science and Letters.
V.~Pietil\"a, T.~P.~Simula, T.~Mizushima, and K.~Machida are appreciated for
useful comments and discussion.
\end{acknowledgments}

\appendix

\section{}

Here, we calculate the dipolar integrals
\begin{equation}
\label{appendix_I} I_\beta(r,\varphi,\kappa)=\sum_\alpha \int d^3r'
D_{\alpha \beta}(\br-\br') \big|_{z=0} M_\alpha(\br')
\end{equation}
appearing in the reduced GP equation, Eq.~(\ref{reduced_GP}). As
shown below, the dipolar potential for a given magnetization can be
efficiently evaluated by carrying out a series of one-dimensional
Hankel transformations.

The expectation values of the spin operators $\hat{S}_\alpha$ are
readily evaluated for the Ansatz in Eq.~(\ref{ansatz}) by applying
the Hadamard lemma. Recalling that quantities evaluated
in the $z=0$ plane are denoted by primes, we obtain
\begin{eqnarray}
\label{appendix_magnetizations} M_x(\br) &=& \cos
(\kappa z) M'_x(r,\varphi+\kappa z) +
\sin (\kappa z) M'_y(r,\varphi+\kappa z),\nonumber \\
M_y(\br) &=& -\sin (\kappa z) M'_x(r,\varphi+\kappa z)
+\cos (\kappa z) M'_y(r,\varphi+\kappa z),\nonumber \\
M_z(\br) &=& M'_z(r,\varphi+\kappa z).
\end{eqnarray}
The components of the planar magnetization are then expanded in
polar Fourier series as
\begin{equation}
\label{Ms}
M'_\gamma(r,\varphi) = \sum_n g_\gamma^n(r) e^{in\varphi},
\end{equation}
where $g_\gamma^n(r)=\int d\varphi M'_\gamma(r,\varphi)
e^{-in\varphi}/2\pi=\left[ g_\gamma^{-n}(r) \right]^*$. For the
helical textures, the components $M'_\gamma$ are in general slowly
varying functions of the azimuthal angle $\varphi$, and hence they
are accurately approximated by only a few terms in the expansion.
The components of magnetization can be written in the form
\begin{equation}
M_\alpha(\br)=\sum_{s,n,\gamma} C_{\alpha \gamma}^s
g_\gamma^n(r) e^{i\left[ s\kappa z + n(\varphi + \kappa z) \right]},
\end{equation}
where $s \in \{1,0,-1\}$ and the coefficients arising from the
trigonometric functions in Eqs.~(\ref{appendix_magnetizations}) are
given by $C^s_{xx}=C^s_{yy}=s^2/2$, $C^0_{zz}=1$,
$C^s_{xy}=-C^s_{yx}=is/2$, and $0$ otherwise.

Each of the three terms in Eq.~(\ref{appendix_I}) is a convolution.
By applying the convolution theorem, we obtain
\begin{equation}
\label{I_equation} I_\beta(r,\varphi,\kappa)=\sum_\alpha
\hat{\mathcal{F}}^{-1} \left\{ K_{\alpha \beta}(\bk)
\hat{\mathcal{F}} M_\alpha(\br)
 \right\}_{z=0},
\end{equation}
where $\hat{\mathcal{F}}$ stands for the Fourier transform. The
functions $K_{\alpha \beta}(\bk)=\hat{\mathcal{F}} D_{\alpha
\beta}(\br)$ can be written in cylindrical coordinates
$(k,\theta,k_z)$ as
\begin{equation}
\label{K_expansion} K_{\alpha \beta}(k,\theta,k_z) = \pi
\sum_{\ell=-2}^2 K_{\alpha \beta}^\ell(k_z/k) e^{i\ell \theta},
\end{equation}
where $K^{\pm 2}_{xx}=-K^{\pm 2}_{yy}=\pm i K^{\pm 2}_{xy}=L(\xi)
\equiv 1/(1+\xi^2)$, $K^{\pm 1}_{xz}=\pm i K^{\pm 1}_{yz}=2\xi
L(\xi)$, $K^0_{xx}=K^0_{yy}=-4/3+2L(\xi)$, $K^0_{zz}=-4/3+4\xi^2
L(\xi)$, $K^\ell_{\alpha \beta}=K^\ell_{\beta \alpha}$, and $0$
otherwise.

The Fourier transforms of the components of magnetization in
Eq.~(\ref{I_equation}) are readily evaluated with the aid of the
Jacobi--Anger expansion, and the result is
\begin{equation}
\label{FS} \hat{\mathcal{F}} M_\alpha(\br)=(2\pi)^2
\sum_{\gamma,s,n} C_{\alpha \gamma}^s (-i)^n H^n_\gamma(k)
e^{in\theta} \delta(\kappa(s+n)-k_z),
\end{equation}
where $H_\gamma^n(k)=\int rdr J_n(kr) g_\gamma^n(r)$ is the $n$th
order Hankel transformation of $g_\gamma^n$. By substituting
Eqs.~(\ref{K_expansion}) and (\ref{FS}) into Eq.~(\ref{I_equation})
and carrying out the integration over $k_z$, we obtain
\begin{eqnarray}
I_\beta(r,\varphi,\kappa)&=&2\pi^2
\sum_{\stackrel{\alpha,\gamma,s}{n,\ell}} C_{\alpha \gamma}^s (-i)^n \\
& &\times \hat{\mathcal{F}}^{-1}_{2D} \left\{ K_{\alpha \beta}^\ell(q)
H_\gamma^n(k) e^{i(n+\ell)\theta}\right\}, \nonumber
\end{eqnarray}
where $q=\kappa(s+n)/k$ and $\hat{\mathcal{F}}^{-1}_{2D}$ denotes
the inverse Fourier transform in the $z=0$ plane. Again, the inverse
Fourier transforms can be evaluated by applying the Jacobi--Anger
expansion, resulting in
\begin{eqnarray}
\label{final_Ibeta} I_\beta(r,\varphi,\kappa)=\pi \sum_{n,\ell}
i^\ell h^{n \ell}_\beta(r) e^{i(n+\ell)\varphi},
\end{eqnarray}
where $h^{n \ell}_\beta(r)=\int k dk J_{n+\ell}(kr) f_\beta^{n
\ell}(k,\kappa)$ stands for the inverse Hankel transformation of
order $n+\ell$ of the function $f_\beta^{n
\ell}(k,\kappa)=\sum_{\alpha,\gamma,s} C_{\alpha \gamma}^s K_{\alpha
\beta}^\ell(q) H_\gamma^n(k)$. By denoting $q_\pm=\kappa (n \pm
1)/k$ and $q_0=\kappa n /k$, these functions can be written as
\begin{eqnarray}
\label{first_f}
f_x^{n, \pm 2} &=& \left( H_x^n \mp i H_y^n \right) L(q_\pm) = \pm
i f_y^{n, \pm 2},\\
f_x^{n, \pm 1} &=& 2 H_z^n q_0 L(q_0) = \pm i f_y^{n, \pm 1},\\
\label{Implicit_Eq} f_x^{n, 0} &=& 2\left( H_x^n \mp i H_y^n
\right)
\left[ -\frac{2}{3} + L(q_\pm) \right] \pm i f_y^{n, 0},
\end{eqnarray}
\begin{eqnarray}
f_z^{n, \pm 1} &=& 2 \left( H_x^n \mp i H_y^n \right) q_\pm
L(q_\pm), \\
\label{last_F} f_z^{n, 0} &=& 2 H_z^n \left[ -\frac{2}{3} + 2 q_0^2
L(q_0) \right],
\end{eqnarray}
where $f_{x}^{n, 0}$ and $f_{y}^{n, 0}$ are given implicitly through
Eq.~(\ref{Implicit_Eq}).

In the case of an axially polarized cylindrically symmetric state,
$\psi_k=\sqrt{n(r)}\delta_{1k}$, the GP equation, Eq.~(\ref{GP}), is
reduced to a scalar equation. In this simple case, $g^0_z$ yields the only
non-vanishing term in Eq.~(\ref{Ms}), which, together with the fact
that $q_0=0$, implies that $f^{0,0}_z=-4 H^0_z/3$ is the only
non-vanishing function in Eqs.~(\ref{first_f})--(\ref{last_F}).
Substitution into Eq.~(\ref{final_Ibeta}) shows that Eq.~(\ref{GP})
is reduced into the scalar equation if we replace $g \longrightarrow
g+g_{\mathrm s}-\frac{4\pi}{3}g_{\mathrm d}$, which is readily
proved also by direct evaluation of Eq.~(\ref{I_equation}).

\bibliography{JabRef}

\end{document}